\documentclass[12pt]{emulateapj}
\def\etal{{\sl et~al.~}}
\newcommand{\eps}{{$\epsilon$~Eri}}
\newcommand{\HST}{{\it HST~}}

\newcommand{\ms}{m s$^{-1}$~}
\def\farcs{\hbox{$.\!\!^{\prime\prime}$}}
\newcommand{\msini}{$M \sin {\it i}~$}

\newcommand{\mjup}{M$_{\rm JUP}~$}

\newcommand{\msun}{M$_{\odot}~$}

\begin{document}
\bibliographystyle{plainnat}
\title{The Extrasolar Planet $\epsilon$ Eridani b - Orbit and Mass\footnote{Based on 
observations made with
the NASA/ESA Hubble Space Telescope, obtained at the Space Telescope
Science Institute, which is operated by the
Association of Universities for Research in Astronomy, Inc., under NASA
contract NAS5-26555} }

\author{ G.\ Fritz Benedict\altaffilmark{2}, Barbara E.
McArthur\altaffilmark{2}, George Gatewood\altaffilmark{5},
 Edmund\ Nelan\altaffilmark{3}, William D. Cochran\altaffilmark{2}, Artie 
Hatzes\altaffilmark{4},  Michael Endl\altaffilmark{2}, Robert 
Wittenmyer\altaffilmark{2}, Sallie L. Baliunas\altaffilmark{6}, Gordon A. H. 
Walker\altaffilmark{7}, Stephenson 
Yang\altaffilmark{7}, Martin K\"{u}rster\altaffilmark{8}, 
Sebastian Els\altaffilmark{9}, and Diane B. Paulson\altaffilmark{10}}

\altaffiltext{2}{McDonald Observatory, University of Texas, Austin, TX 78712}
\altaffiltext{3}{Space Telescope Science Institute, 3700 San Martin Dr., 
Baltimore, MD 
21218}
\altaffiltext{4}{Tautenburg Observatory, Germany}
\altaffiltext{5}{Allegheny Observatory, University of Pittsburgh, Pittsburgh, 
PA}
\altaffiltext{6}{Harvard-Smithsonian Center for Astrophysics, Cambridge, MA 
02138}
\altaffiltext{7} {Physics \& Astronomy Dept., University of Victoria, BC, V8W 3P6, Canada}
\altaffiltext{8}{Max-Planck-Institut f\"ur Astronomie, K\"onigstuhl 17, Heidelberg,
D-69117. Germany.}
\altaffiltext{9} {Cerro Tololo Interamerican Observatory, La Serena, Chile.}
\altaffiltext{10} {Goddard Spaceflight Center, Greenbelt, MD 20771.}
% Notice that each of these authors has alternate affiliations, which
% are identified by the \altaffilmark after each name.  The actual alternate
% affiliation information is typeset in footnotes at the bottom of the
% first page, and the text itself is specified in \altaffiltext commands.
% There is a separate \altaffiltext for each alternate affiliation
% indicated above.

% The abstract environment prints out the receipt and acceptance dates
% if they are relevant for the journal style.  For the aasms style, they
% will print out as horizontal rules for the editorial staff to type
% on, so long as the author does not include \received and \accepted
% commands.  This should not be done, since \received and \accepted dates
% are not known to the author.

\begin{abstract}
{\it Hubble Space Telescope} (\HST) observations of the nearby (3.22 pc), K2 V star 
$\epsilon$ Eridani  have been combined with ground-based astrometric and radial 
velocity  data to determine the mass of its known companion. We model the astrometric 
and radial velocity measurements simultaneously to obtain the parallax, proper 
motion, perturbation period, perturbation inclination, and perturbation size.  Because 
of the long period of the companion, \eps~b, we extend our astrometric coverage to a total of 14.94 years (including the three year span of the \HST data) by including lower-precision ground-based astrometry from the Allegheny Multichannel Astrometric Photometer.  Radial velocities now span 1980.8 -- 2006.3. We obtain a perturbation period,   P = 6.85 $\pm$ 0.03 yr,  semi-major axis $\alpha =1.88 \pm 0.20$ mas, and  
inclination $i$ = 30\fdg1 $\pm$ 3\fdg8. This inclination is consistent with a 
previously measured dust disk inclination, suggesting coplanarity.  Assuming a primary mass $M_* = 0.83 M_{\sun}$, we obtain a companion mass {\it M} = 1.55 $\pm$ 0.24 {\it 
M}$_{Jup}$. Given the relatively young age of $\epsilon$~Eri ($\sim$800 Myr), 
this accurate exoplanet mass and orbit can usefully inform future direct imaging 
attempts. We predict the next periastron at 2007.3 with a total separation, $\rho$ = 0\farcs3 at position angle, p.a. = -27\arcdeg. Orbit orientation and geometry dictate that \eps~b will appear brightest in reflected light very nearly at periastron. Radial velocities spanning over 25 years indicate an acceleration consistent with a Jupiter-mass object with a period in excess of 50 years, possibly responsible for one feature of the dust morphology, the inner cavity.
\end{abstract}

% The different journals have different requirements for keywords.  The
% keywords.apj file, found on aas.org in the pubs/aastex-misc directory,
% contains a list of keywords used with the ApJ and Letters.  These are
% usually assigned by the editor, but authors may include them in their
% manuscripts if they wish.


% That's it for the front matter.  On to the main body of the paper.
% We'll only put in tutorial remarks at the beginning of each section
% so you can see entire sections together.

% In the first two sections, you should notice the use of the LaTeX \cite
% command to identify citations.  The citations are tied to the
% reference list via symbolic KEYs.  We have chosen the first three
% characters of the first author's name plus the last two numeral of the
% year of publication.  The corresponding reference has a \bibitem
% command in the reference list below.
%
% Please see the AASTeX manual for a more complete discussion on how 
%to make \cite-\bibitem work for you.

\section{Introduction}

$\epsilon$ Eridani ( = HD 22049 = HIP 16537 =  HR 1084 = PLX 742), with a 
spectral 
type of K2V,  is one of the  nearest solar-type  stars with a distance of about 3.2 pc. 
It is slightly metal-poor (Fe/H = -0.13$\pm$ 0.04, Santos \etal 
2004\nocite{San04}, Laws \etal 2003).
Its proximity makes it a prime target for future extrasolar planet direct imaging efforts. The success of these efforts will depend on knowing exactly where to look, requiring accurate orbital elements for the companion. It will depend on the mass of the planetary companion, and will depend on the age of the system. Younger and more massive gas giant planets are predicted to be brighter (Hubbard \etal 2002). If young enough, the intrinsic luminosity of  \eps ~b  might be greater than its  brightness in reflected light. However, planetary mass objects with the age of \eps~ 
change intrinsic luminosity by a factor of nearly 100 between  1 $M_{Jup}$ and 7 
$M_{Jup}$. Hence the need for a more precise companion mass.

\eps~has  
been the subject of multiple radial velocity (RV) planet searches. Walker  \etal  
(1995), using measurements spanning 11 years, found evidence
for $\approx$ 10 yr variation with an amplitude of 15 {\ms}. These
results were substantiated by Nelson \& Angel (1998) using an analysis
of the same data set. Cumming  \etal  (1999)
analyzed 11 years of RV data on this star taken at Lick Observatory and
found significant variations with comparable amplitude  but with a shorter 
period  of 6.9 years. Because of the high
level of magnetic activity for $\epsilon$~Eri (inferred from chromospheric activity),
these RV variations were largely interpreted as arising
from a stellar activity cycle.  The McDonald Observatory Planet Search Program 
(Cochran \& Hatzes 1999) has
monitored $\epsilon$~Eri  since late 1988.
The McDonald results in combination with these other
surveys, along with data from ESO (Endl \etal 2002), confirmed the presence of 
long period RV variations
and demonstrated that the most likely explanation for the observed RV variations 
was  the presence of a planetary companion with a period, P = 6.9 y. Details of this analysis are given in
Hatzes  \etal  (2000).

We obtained three years of astrometry with {\it HST} with millisecond of arc precision  which we  combined with  radial velocity data as we 
have in previous planetary mass studies \cite{Ben02b,McA04}. Just as in the case 
of the Hatzes \etal (2000) radial velocity analysis, where less-precise data extended 
the observation span and allowed a companion detection, we anticipated that 
less-precise astrometry with a 14 year baseline from the Allegheny Multichannel Astrometric 
Photometer (MAP) astrometry project would improve the astrometric result.
Gatewood (2000) reported in a meeting abstract an inclination, i = 
46\arcdeg $\pm$ 
17\arcdeg ~and a companion mass $M_b = 1.2 \pm 0.3 M_{Jup}$ obtained with the MAP data alone. In this study we combine the MAP data with the \HST data, only to improve the determination of the proper motion of \eps.  The parallax and proper motion must be removed as accurately as possible to determine the perturbation orbit of \eps, which, when combined with an estimate of the mass of \eps, will provide  the mass of the companion, \eps~b.

This paper presents a mass of the planet orbiting \eps~ discussed in Hatzes \etal (2000), not the far longer period object inferred from dust-disk morphology \citep{QT02}. Our mass is derived from 
combined astrometric and radial velocity data, continuing a series 
presenting accurate masses of planetary companions to nearby 
stars. Previous results include the mass of Gl 876b \citep{Ben02b} and of 
$\rho^1$ Cancri d \citep{McA04}.

In Section 2 we briefly review  the astrometers and discuss the data sets coming from each, and identify our many sources for radial velocities. In Section 3 we present the results of extensive spectrophotometry of the astrometric reference stars, information required to correct relative parallax to absolute. In Section 4 we briefly discuss our astrometric modeling and the quality of our results as determined by residuals. In Section 5 we review our radial velocity data. In Section 6 we derive an absolute parallax and relative proper motion for \eps, those nuisance parameters that must be removed to determine the perturbation orbital parameters.  We finally establish the perturbation orbital parameters and, combined with an estimate of the mass of \eps, estimate a mass for \eps ~ b. We discuss system age, dust, and companion detectability in Section 7, and  summarize our conclusions in Section 8.

\section{The Astrometers and Observational Data}

\subsection{\HST  FGS1r}
We used {\it HST} Fine Guidance Sensor 1r (FGS1r) to carry out our space-based  
astrometric observations. Nelan \etal (2003)\nocite{Nel03}  provides a detailed 
overview of FGS1r as a science instrument. Benedict \etal (2002b) \nocite{Ben02b} 
describe the FGS3 instrument's astrometric capabilities along with the data acquisition and 
reduction strategies used in the present study. We use FGS1r for the present study because 
it provides superior fringes from which to obtain target and reference star 
positions  \citep{McA02}.

Table~\ref{tbl-LOOF} presents a log of \HST FGS observations. Epochs 2--4 
contain multiple data sets acquired contiguously, the time span less than a day. Each 
time is that of the first observation within each epoch. Each distinct observation set 
typically contains five measurements of \eps. The field was observed at multiple spacecraft roll 
values, and \eps ~had to be placed in different, non-central locations within the FGS1r FOV 
to accommodate the less than optimal distribution of reference stars. FGS 
photometric sensitivity depends on location within the FOV ({\it e.g.} Benedict \etal 1998) 
and depends on time, as the FGS PMT age. Given the faintness of our reference stars, 
we could not use them to provide a high-precision flat field. Hence,   we could not 
extract millimagnitude photometry  ({\it c.f.} Benedict \etal 2000a) to monitor \eps~ 
stellar activity.

\subsection{The Allegheny MAP}
The Multichannel Astrometric Photometer (MAP) and associated observation and
reduction procedures are described by Gatewood (1987).  The observational program
utilizing the MAP began in 1986, but reluctance to observe at the low \eps~
declination over the city of Pittsburgh delayed initiation of its observation
until January of 1989.  Despite the reduced precision and rate of successful
observation, the field remained on the MAP program until the installation of new
instrumentation early in 2004 (Gatewood 2004).

MAP observations of the brightest stars utilize either a specially filtered 12th
channel (e.g. Gatewood \& Han 2006) or a divide by 16 feature on channel 2, which reduces
the count sufficiently for the 16 bit counters.  With an R magnitude of 3.0  \eps~ could be placed on channel 2 without counter flooding.  The other 10
channels were assigned to the MAP reference stars noted below.  An observation
consists of 4 x 11 minute sweeps of the ruling across the field with probe
rotations and ruling rotations to reduce systematic error (Gatewood 1987). Thus
each observation consists of approximately 22 minutes of integration on each
axis.  
Table~\ref{tbl-LOOM} presents a log of MAP observations.

\subsection{Radial Velocities}
The radial velocity data used include all data described in Hatzes \etal 
(2000)\nocite{Hat00}, in addition to more recent data from McDonald Observatory 
and ESO.  These data now span over 25 years. All sources are listed in Table~\ref{tbl-RV}. Briefly, McDonald Phases I, II, and III are all data obtained with the 2.7m Smith telescope. The phases correspond (I) to early velocities referenced to atmospheric O$_2$, (II) velocities obtained with an I$_2$ cell, and (III) velocities obtained with the McDonald 2d-Coude spectrograph (Tull \etal 1995) and an I$_2$ cell.  The only new radial velocity data included in this new study are from the McDonald Observatory Phase III 2.7m program. They are listed in Table~\ref{tbl-P3}. Note that the errors associated with these data are larger than typically produced by this telescope/spectrograph combination \citep{End06} due to high levels of~\eps ~stellar activity.

\section{\eps  ~Astrometric Reference Frames}  \label{AstRefs}
Any prior knowledge concerning the fifteen stars included in  our reference frame 
eventually enters our modeling as observations with error, and yields the most 
accurate parallax and proper motion for the prime target, \eps. These periodic and non-
periodic motions must be removed as accurately and precisely as possible to obtain the 
perturbation inclination and size caused by \eps~b. 

\subsection{The MAP Reference Frame}
\begin{center}
\begin{figure}
\epsscale{0.65}
\plotone{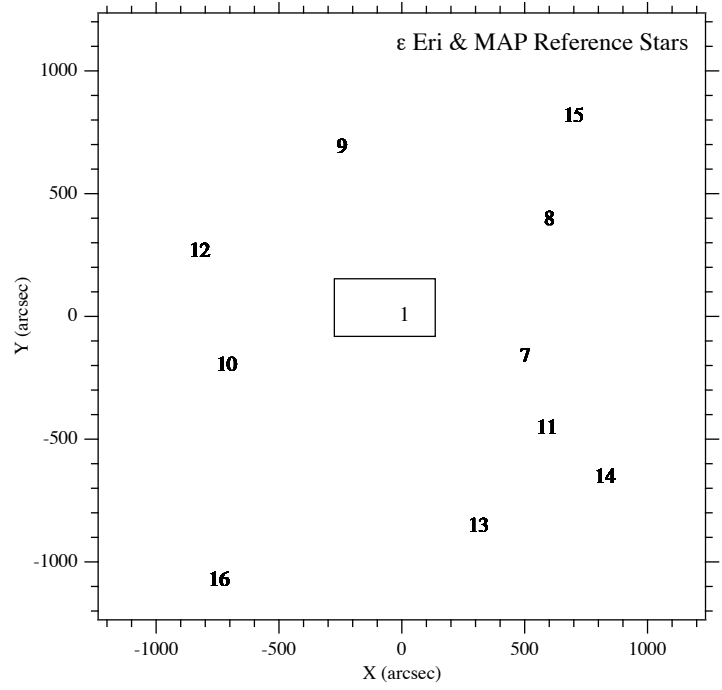}
\caption{\eps ~and MAP reference
stars on the sky. Each star is identified by the number listed in 
Table~\ref{tbl-2MASSPos}. The inner box indicates the FGS1r reference frame coverage for epoch 2 in Table~\ref{tbl-LOOF}. }
\label{fig-1}
\end{figure}
\end{center}

Figure \ref{fig-1} shows the distribution  of the ten reference stars in the 130 
MAP  \eps ~observation sets (Table~\ref{tbl-LOOM}). Note that the areal coverage is 
approximately 0\fdg6$\times$0\fdg6, allowing for the use of relatively  bright 
reference stars, well-distributed around the prime target \eps, in contrast to the case for the \HST FGS (below).

\subsection{The FGS Reference Frame}
\begin{center}
\begin{figure}
\epsscale{0.65}
\plotone{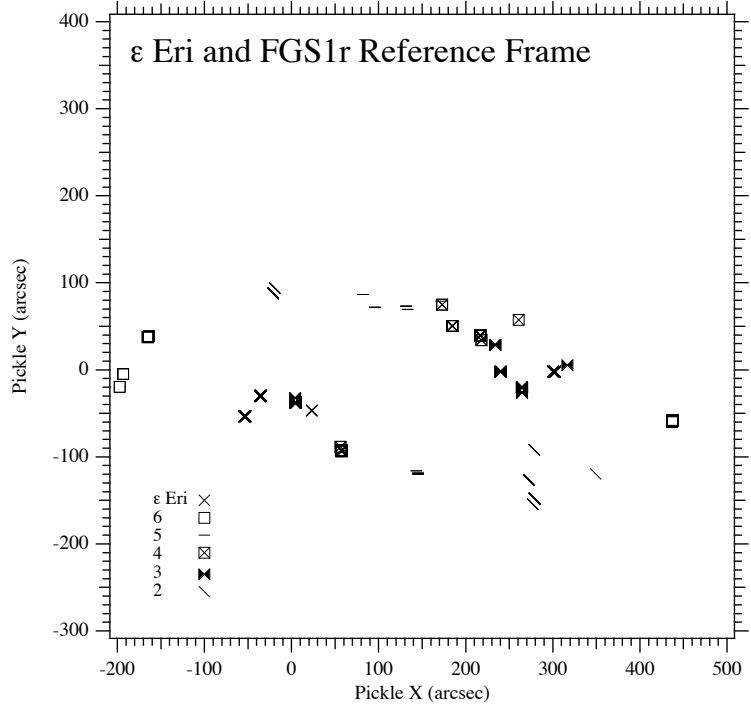}
\caption{ \HST ~\eps ~and reference
frame observations in FGS1r pickle coordinates. The symbol shape
identifies each star listed in Table~\ref{tbl-2MASSPos}. Note that the position 
of \eps 
~($\times$) within the FGS1r FOV is not fixed at the center.}
\label{fig-2}
\end{figure}
\end{center}

Figure \ref{fig-2} shows the distribution in FGS1r pickle coordinates of the 52
 sets of five reference star measurements for the \eps ~ field. The arcing 
pattern is enforced by the requirement that {\it HST} must roll to keep its solar panels fully 
illuminated throughout the year. To ensure access to all reference stars for every observation set,  it was not possible to keep \eps~ (symbol, 
$\times$) located in the center of the FGS1r FOV. At each epoch we measured each reference stars 2 -- 4 
times, and \eps  ~4--5 times.

\subsection{Absolute Parallaxes for the Reference Stars}
Because the parallax determined for \eps  ~ is measured with respect to 
reference frame stars which have their own parallaxes, we must either apply a statistically-
derived correction from relative to absolute parallax \cite[Yale Parallax Catalog, YPC95]{WvA95}, or 
estimate the absolute parallaxes of the reference frame stars.  In principle, 
the colors, spectral type, and luminosity class of a star can be used to estimate the 
absolute magnitude, $M_V$, and $V$-band absorption, $A_V$. The absolute parallax is then 
simply,
\begin{equation}
\pi_{\rm abs} = 10^{-(V-M_V+5-A_V)/5}
\end{equation}

\subsubsection{ Reference Star Photometry}
\begin{center}
\begin{figure}
\epsscale{0.65}
\plotone{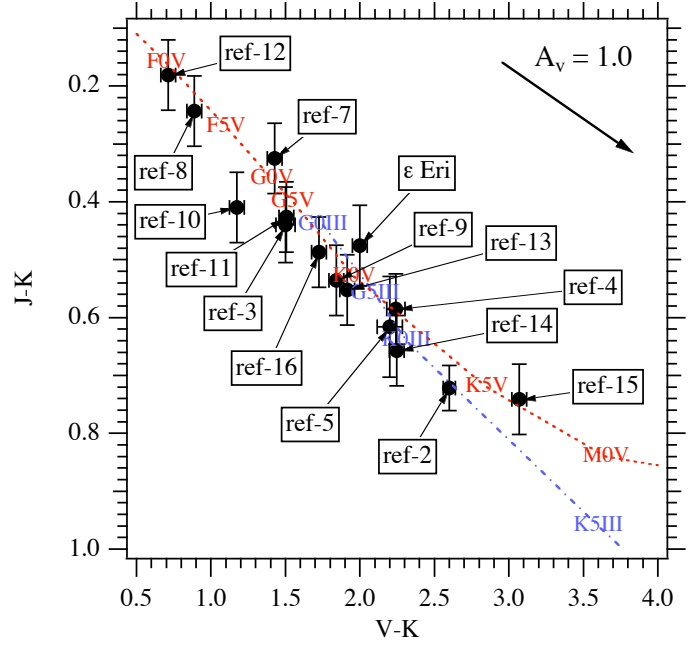}
\caption{$(J-K)$ vs. $(V-K)$ color-color diagram for stars identified in 
Table~\ref{tbl-2MASSPos}. The dashed line is the locus of  dwarf
(luminosity class V) stars of various spectral types; the dot-dashed line is for 
giants (luminosity class III). The reddening vector indicates $A_V = 1.0$ for the 
plotted color systems. Along this line of sight maximum extinction is  $A_V$$\sim$ 0.1 
\citep{Sch98}. }
\label{fig-3}
\end{figure}
\end{center}

Our bandpasses for reference star photometry include:  $V$ (from FGS1r), and 
$JHK$ from 2MASS\footnote{The Two Micron All Sky Survey is a joint project of the 
University of Massachusetts and the Infrared Processing and Analysis 
Center/California Institute of Technology }.  The  $JHK$ values have been transformed to the 
Bessell \& Brett (1988)\nocite{Bes88} system using the transformations provided in 
Carpenter (2001)\nocite{Car01}. Table \ref{tbl-IR} lists $VJHK$ photometry for the target 
and reference stars indicated in Figures \ref{fig-1} and \ref{fig-2}. Figure 
\ref{fig-3} contains a $(J-K)$ vs. $(V-K)$ color-color diagram with  reference stars and 
\eps ~ labeled. Schlegel \etal (1998)  find an upper limit $A_V$$\sim$ 0.1 towards 
\eps. In the following we adopt  $\langle A_V \rangle$ = 0.0, but increase the error on 
reference star distance moduli by 0.1 magnitudes to account for absorption uncertainty.

\begin{center}
\begin{figure}
\epsscale{0.65}
\plotone{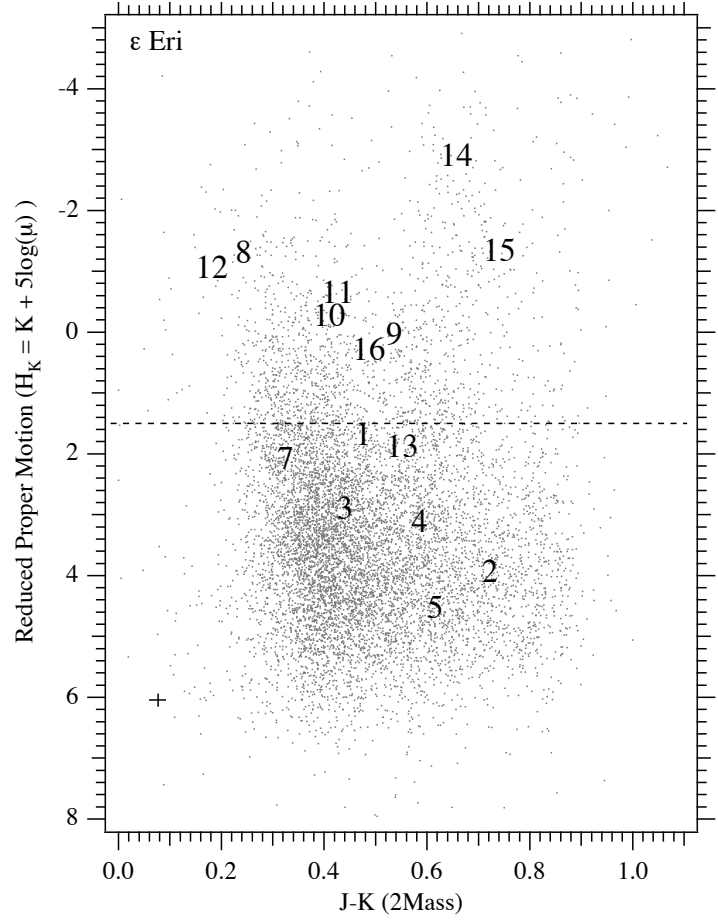}
\caption{Reduced proper motion diagram for 9,041 stars in a 6\arcdeg ~field centered
on \eps. Star identifications are in Table~\ref{tbl-2MASSPos}. For a given spectral type, 
giants and sub-giants have more negative $H_K$ values and are redder than dwarfs in 
$(J-K)$.  $H_K$ values are derived from `Final' proper motions in Table \ref{tbl-PM}. The small cross at the lower left represents a typical $(J-K)$ error of 0.04 mag and $H_K$ error of 0.17 mag. The horizontal dashed line is a giant-dwarf demarcation derived from a statistical analysis of the Tycho input catalog (Ciardi 2004, private communication). Ref-14 and -15 are likely luminosity class III.} 
\label{fig-4}
\end{figure}
\end{center}

The derived absolute magnitudes are critically dependent on the assumed stellar 
luminosity, a parameter impossible to obtain for all but the latest type stars 
using only  Figure \ref{fig-3}. To confirm the luminosity classes we  obtain  UCAC2 proper 
motions \citep{Zac04} for a one-degree-square field centered on \eps , and then 
iteratively employ the technique of reduced proper motion \citep{Yon03,Gou03}  to discriminate 
between giants and dwarfs. The end result of this process is contained in 
Figure~\ref{fig-4}.

\subsubsection{Adopted Reference Frame Absolute Parallaxes}\label{CORR}

We derive absolute parallaxes using our estimated spectral types and luminosity 
class and $M_V$ values from \nocite{Cox00} Cox (2000). Our adopted input errors for 
distance moduli, $(m-M)_0$, are 0.4 mag for all reference stars (except ref-2 and -6, as 
discussed below). Contributions to the error are a small but undetermined $A_V$ and errors 
in $M_V$ due to uncertainties in color to spectral type mapping. We estimate a 
spectral type for reference star ref-6 only through its apparent magnitude, hence, the 
larger error in its distance modulus. Ref-2, which in Figure~\ref{fig-3} straddles the gap 
between giants and dwarfs,  was finally typed K4V, because the $\chi^2$ (from modeling of the reference 
frame) significantly decreased, using that typing.  Its input parallax error was 
also increased. All reference star absolute parallax estimates are listed in Table \ref{tbl-SPP}.  Individually, no reference star absolute parallax is better determined than ${\sigma_{\pi}\over \pi}$ = 
18\%. The average input absolute parallax for the reference frame is 
$\langle\pi_{abs}\rangle = 4.9$ mas, a quantity known to $\sim5$\% (standard deviation of the mean of fifteen reference 
stars). We compare this to the correction to absolute parallax discussed and presented in 
YPC95 (Sec. 3.2, Fig. 2).  Entering YPC95, Fig. 2, with the Galactic latitude of \eps, $b = 
-48\arcdeg$, and average magnitude for the reference frame, $\langle V_{\rm ref} \rangle 
= 12.2$, we obtain a correction to absolute of 2.3 mas, considerably different. Rather than 
apply a model-dependent correction to absolute parallax, we introduce our 
spectrophotometrically-estimated reference star parallaxes  into our reduction 
model as observations with error.

\section{The Astrometric Model}
The \eps ~ reference frame contains fifteen stars and has been measured by two different 
astrometers, FGS1r and MAP. The only object in common is \eps. From these 
positional measurements we determine the scale, rotation, and offset ``plate
constants" relative to an arbitrarily adopted constraint epoch for
each observation set. As for all our previous astrometric analyses, we employ GaussFit (Jefferys \etal 
1988) 
\nocite{Jef88} to minimize $\chi^2$. The solved equations of condition for the 
\eps ~ 
field are:
\begin{equation}
        x^\prime = x + lcx(\it B-V)  - \Delta XFx
\end{equation}
\begin{equation}
        y^\prime = y + lcy(\it B-V) - \Delta XFy
\end{equation}
\begin{equation}
\xi = Ax^\prime + By' + C  - \mu_\alpha \Delta t  - P_\alpha\pi
\end{equation}
\begin{equation}
\eta = Dx^\prime + Ey^\prime + F  - \mu_\delta \Delta t  - P_\delta\pi
\end{equation}

\noindent for FGS1r data
and
\begin{equation}
\xi =Ax + By + C-P_\alpha\pi - \mu_\alpha  \Delta t 
\end{equation}
\begin{equation}
\eta = Dx +Ey +F-P_\delta\pi - \mu_\delta  \Delta t 
\end{equation}
\noindent for the MAP data.
Identifying terms, $\it x$ and $\it y$ are the measured coordinates from {\it HST} and the MAP;   $(B-V)$ represents the $(B-V)$ color of each star, estimated from its spectral type, $A_V$, and $(J-K)$ color listed in Table~\ref{tbl-IR}; and $\it lcx$ and $\it lcy$ are the lateral color corrections, applied only to FGS1r data. Here $\Delta$XFx and $\Delta$XFy are the cross filter corrections in $\it x$ and $\it y$, applied to the observations of \eps~in  FGS1r.  $A$, $B$, $D$ and $E$ are scale- and rotation plate constants, $C$ and $F$ are offsets; $\mu_\alpha$ and $\mu_\delta$ are proper motions; $\Delta t$ is the epoch difference from the mean epoch; $P_\alpha$ and $P_\delta$ are parallax factors;  and $\it \pi$ is  the parallax.   We obtain the parallax factors from a JPL Earth orbit predictor 
(Standish 1990)\nocite{Sta90}, upgraded to version DE405. Orientation to the sky for the 
FGS1r data is obtained from ground-based astrometry (2MASS Catalog) with uncertainties 
of $0\fdg01$.

\subsection{Assessing Reference Frame Residuals}
\begin{figure}
\epsscale{0.65}
\plotone{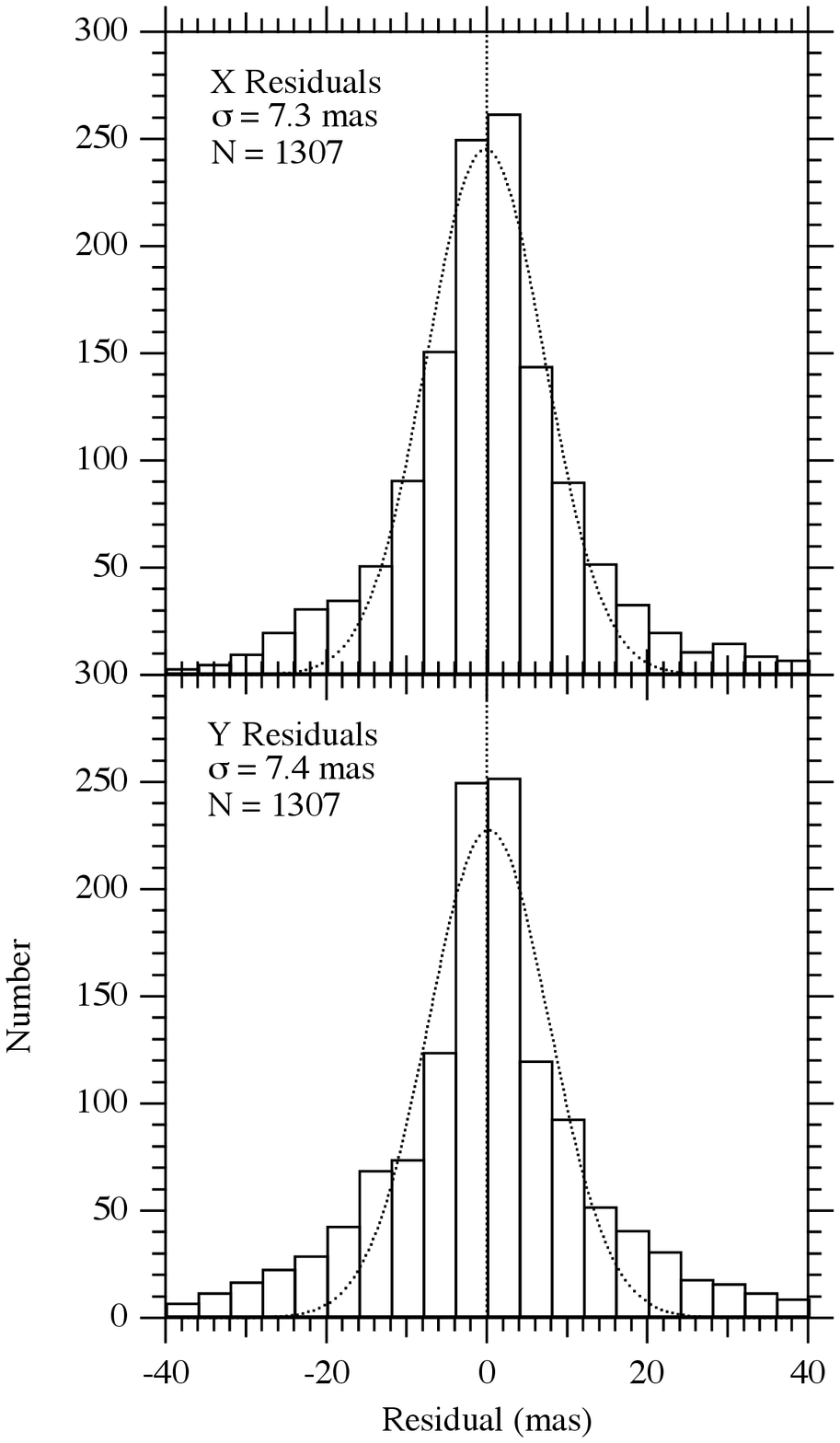}
\caption{Histograms of x and y residuals obtained from modeling the MAP observations 
of \eps ~and the MAP reference frame with equations 6 and 7. Distributions are fit with Gaussian distributions.} \label{fig-MAPH}
\end{figure}

\begin{figure}
\epsscale{0.65}
\plotone{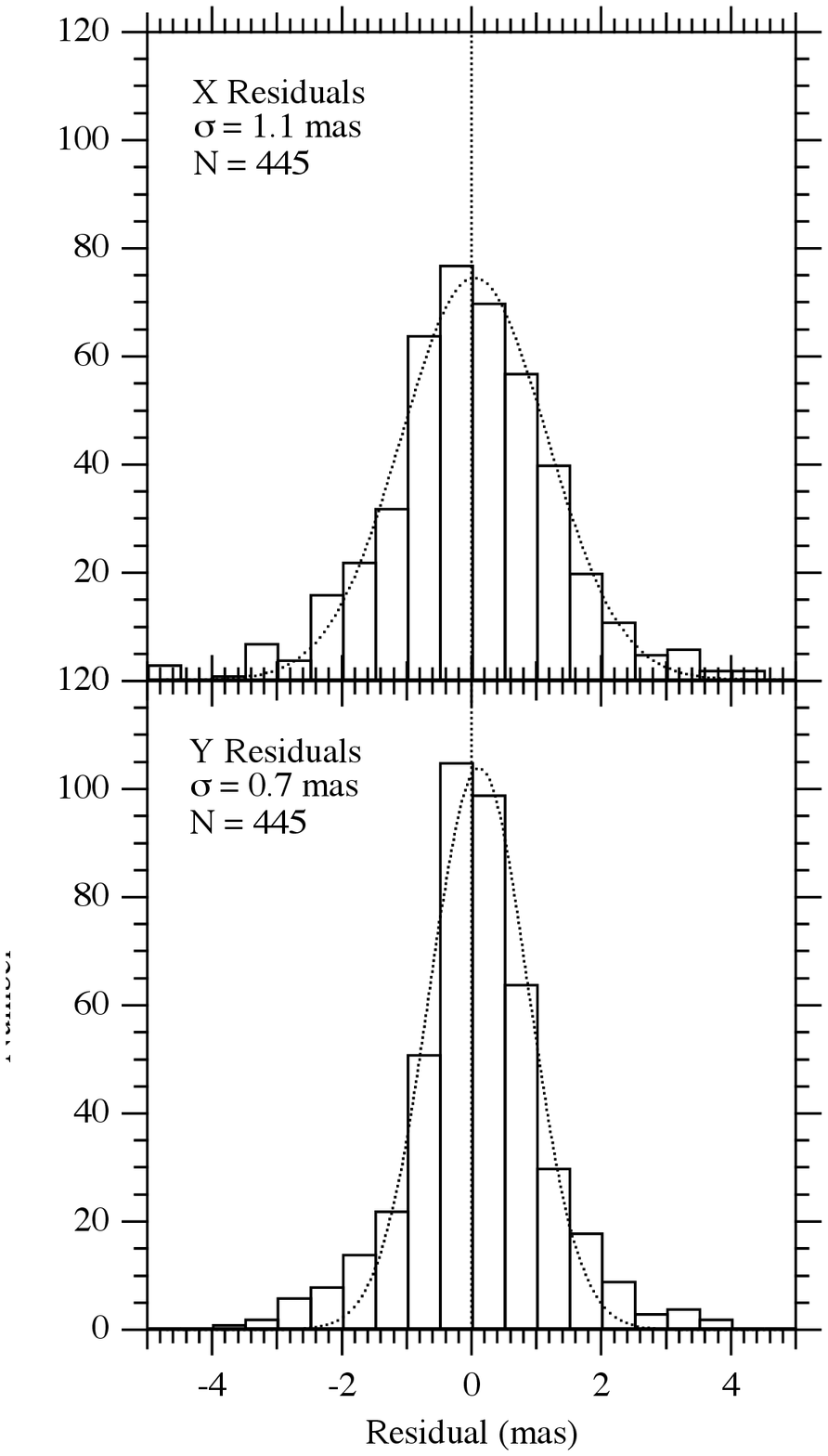}
\caption{Histograms of x and y residuals obtained from modeling the FGS 
observations of \eps ~and the FGS reference frame with equations 4 and 5. Distributions are 
fit with Gaussian distributionss.} \label{fig-FGSH}
\end{figure}

Histograms of the MAP residuals (Figure~\ref{fig-MAPH}) indicate 
per-observation precision  of $\sim$7 mas. Because we are seeking the signature of a perturbation over three times smaller than that per-observation precision, the MAP data were only used to lower the errors on parallax and proper motion, not to establish any perturbation parameters. As for the FGS data, the Optical Field Angle Distortion calibration \cite{McA02} 
reduces as-built {\it HST} telescope and FGS1r distortions with magnitude 
$\sim1\arcsec$ to below 2 mas  over much of the FGS1r field of regard. From 
histograms of the FGS astrometric residuals (Figure~\ref{fig-FGSH}) we conclude 
that we have obtained correction at the $\sim 1$ mas level. The  reference frame 
'catalogs' for MAP and FGS1r in $\xi$ and $\eta$ standard coordinates (Table \ref{tbl-POS}) 
were determined with $<\sigma_\xi>= 1.0$  and $<\sigma_\eta> = 1.3$ mas (MAP), and  
$<\sigma_\xi>= 0.3$	 and	$<\sigma_\eta> = 0.2$ mas (FGS). 

%To determine if there might be unmodeled - but possibly correctable -  systematic effects at the 1 mas level, we plotted the \eps~reference frame X and Y residuals against a number of spacecraft, instrumental, and astronomical parameters. These included X, Y position within the pickle; radial distance from the pickle center; reference star V magnitude and B-V color; and epoch of observation.  We saw no obvious trends, other than an expected increase in positional uncertainty with reference star magnitude. 

\section{Radial Velocities}
Measurements from four planet search groups were included in our modeling. 
Table~\ref{tbl-RV} lists the source, coverage, technique, number of observations 
and the rms deviation from the final orbit of these observations. The weighting of the RV data was carefully 
evaluated with independent modeling and significant outliers were filtered. For 
example - if five data points were taken in succession, all assigned with the 
same weight and one point was 50 \ms offset from the others, that point 
was discarded for this solution.   Initially these points were merely 
re-weighted, but later discarded as spurious data.  The total number of observations so discarded was less than 1\% of the aggregate data.  This improved our goodness of fit ($\chi^2$ /degrees of freedom)  measurement of the modelling for the RV data set from 0.94 in the announcement paper to 0.30 in the current analysis.

\section{ \eps~ Parallax, Proper Motion, and Perturbation Orbit from Astrometry 
and 
Radial Velocities}

Solving for relative parallax, proper motion, and orbital motion, using 
astrometry and radial velocities  simultaneously, the model now becomes,

\begin{equation}
\xi = aX + bY + c - P_x*\pi - \mu_x*t - ORBIT_x
\end{equation}
\begin{equation}
\eta = -bX +aY +f - P_y*\pi - \mu_y*t - ORBIT_y
\end{equation} 

\noindent where ORBIT is a function (through Thiele-Innes constants) of the 
traditional astrometric and radial velocity orbital elements listed in Table~\ref{tbl-ORB}.

The period (P), the epoch of passage through periastron in years (T), the
eccentricity (e) and the angle in the plane of the true orbit between
the line of nodes and the major axis ($\omega$), are constrained to be equal
for the radial velocity and astrometry portions of the model.  Only radial 
velocity provides information with which to determine the half-amplitudes ($K_1$) and 
$\gamma$, the systemic velocity.  Combining radial velocity observations from different sources is possible with GaussFit, which has the ability to simultaneously solve for many separate velocity offsets (because velocities from different sources are relative, having differing  zero points), along with the other orbital parameters. 

We force a relationship between the astrometry and the radial velocity 
 by a constraint from \cite{Pou00}
\begin{equation}
\displaystyle{{\alpha_A~sin~i \over \pi_{abs}} = {P K_1 (1 -
e^2)^{1/2}\over2\pi\times4.7405}} 
\end{equation}
\noindent where quantities derived only from astrometry (parallax, $\pi_{abs}$, 
primary perturbation orbit size, $\alpha_A$, and inclination, $i$) are on the left, and 
quantities derivable from both (the period, $P$ and eccentricity, $e$), or radial 
velocities only (the radial velocity amplitude for the primary, $K_1$), are on the right.

Combining RV measurements complete through 2006.3 (Table~\ref{tbl-RV}), all the 
astrometric measurements, and the Equation 10 constraint, we solve for parallax, 
proper motion, and the semi-major axis, orbit orientation, and orbit inclination for 
the perturbation caused by the companion. For the parameters critical in determining 
the mass of \eps ~we find a parallax, $\pi_{abs} = 311.37 \pm 0.10$ mas and a proper 
motion  $976.54\pm 0.1$  mas y$^{-1}$ in position angle 269\fdg0  $\pm$ 0\fdg6. 
Table~\ref{tbl-SUM}  compares values for the parallax and proper motion of \eps 
~from {\it HST} and {\it HIPPARCOS}. We note satisfactory agreement. Our precision and 
extended study duration have significantly improved the accuracy and precision of the 
parallax and proper motion of \eps.

\begin{figure}
\epsscale{0.65}
\plotone{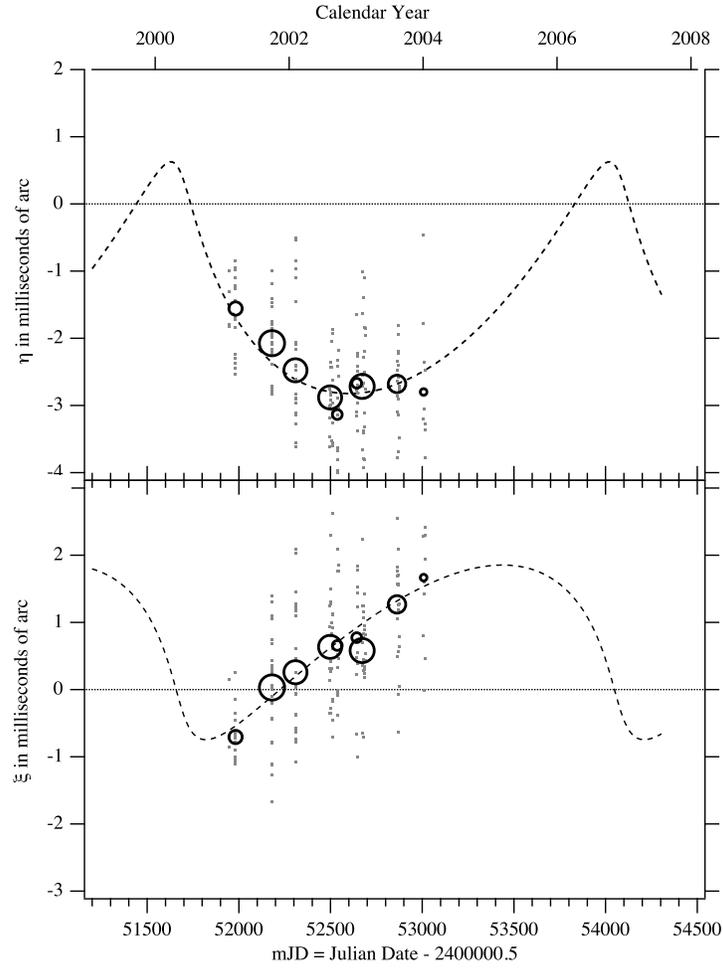}
\caption{X and Y components of the perturbation orbit for \eps ~as a function of time.   
The dashed line is the orbit described by the orbital elements found in Table~\ref{tbl-ORB}. The dots are all the \eps~ observation residuals to a model that does not 
contain orbital motion. The circles are normal points formed from the individual 
observation residuals, where the circle size is proportional to the number of observations 
forming the normal point. } \label{fig-OrbXY}
\end{figure}

At this stage we can assess the reality of any \eps~perturbation by plotting 
residuals to a model that does not include an orbit. Figure~\ref{fig-OrbXY} shows the X and Y 
components of only the higher-precision astrometry FGS residuals plotted as grey 
dots. The lower precision MAP data were not considered in  the determination of 
the orbital parameters. We also plot normal points formed from those dots at nine epochs. Finally, each 
plot contains as a dashed line the X and Y components of the perturbation we find by 
including an orbit in our modeling.

\begin{center}
\begin{figure}
\epsscale{1.00}
\plotone{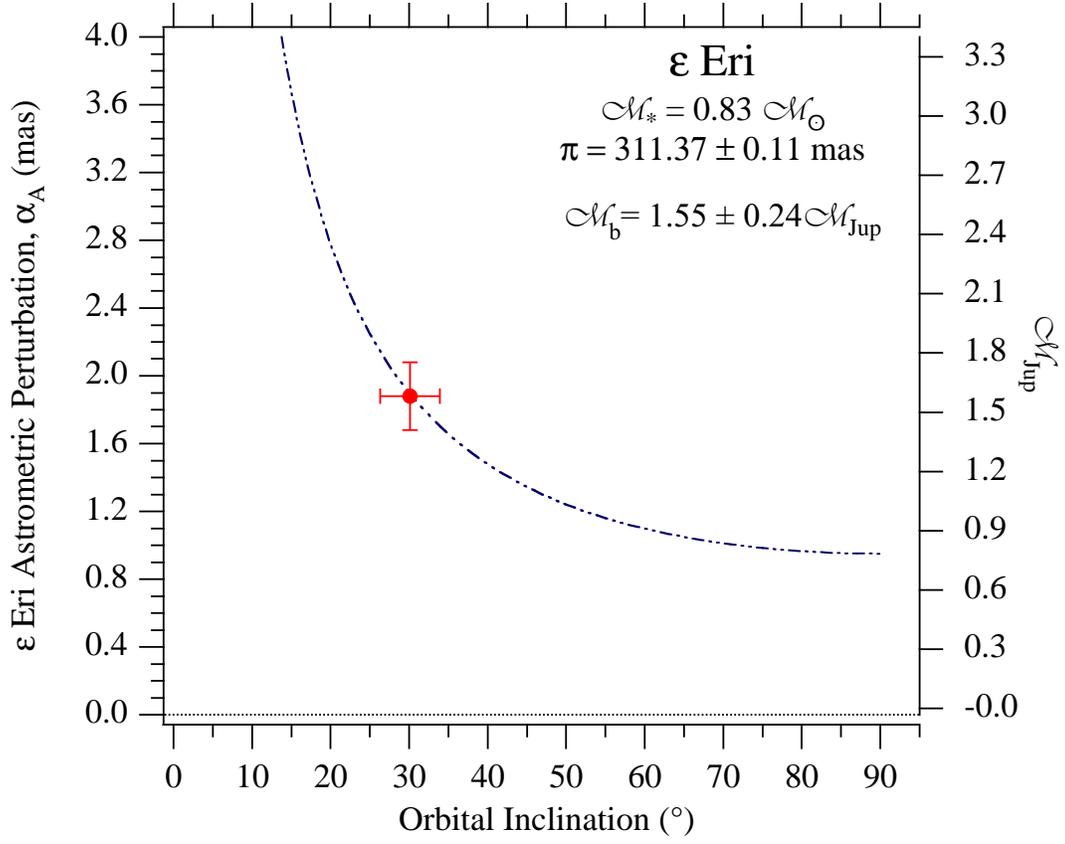}
\caption{This curve relates perturbation size and inclination for the \eps ~perturbation through the Pourbaix \& Jorrisen (2000)  relation (Equation 10). We use the curve as a 'prior' in a quasi-Bayesian sense. Our final values for the semimajor axis of the astrometric perturbation, $\alpha_A$, and inclination, $i$ are plotted with the formal errors.} 
\label{fig-PJconsTEMP}
\end{figure}
\end{center}

\begin{figure}
\epsscale{1.0}
\plotone{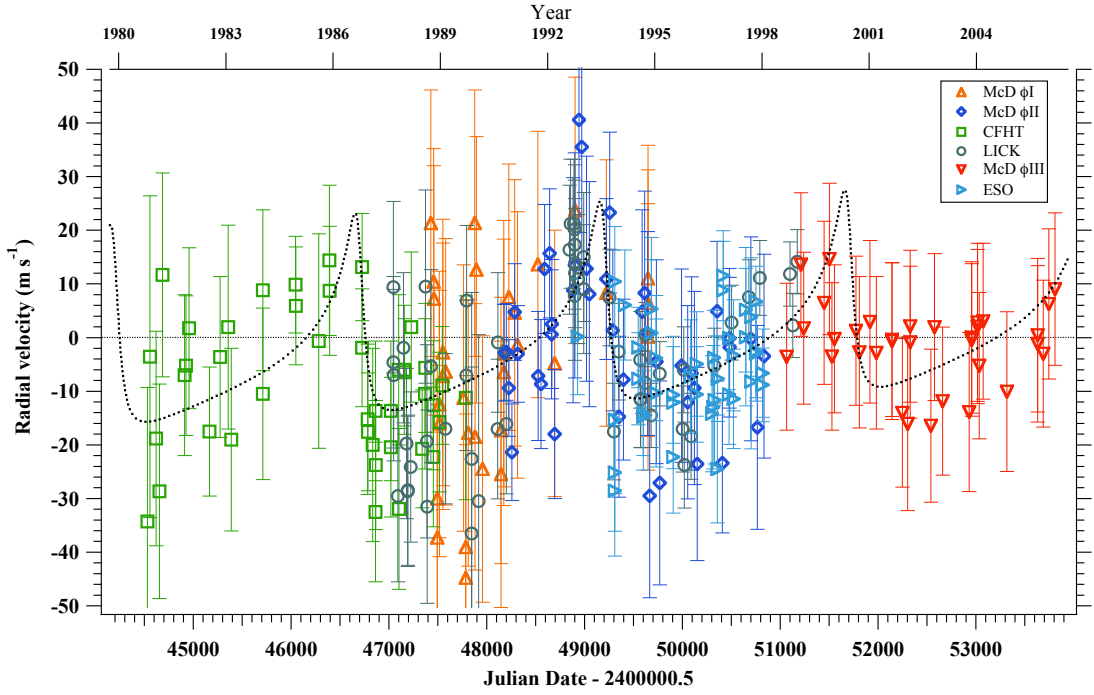}
\caption{Radial velocity measurements of \eps ~from sources as indicated in the legend 
(and identified in Table~\ref{tbl-RV}).  The 
dashed line is the velocity predicted from the orbital parameters (Table~\ref{tbl-ORB}) 
derived in the combined astrometry and radial velocity solution (Section 6). The linear trend in velocity can be seen by comparing the minima in the radial velocity orbit near 1981 with that at 2001.} \label{fig-RVS}
\end{figure}

\begin{figure}
\epsscale{1.0}
\plotone{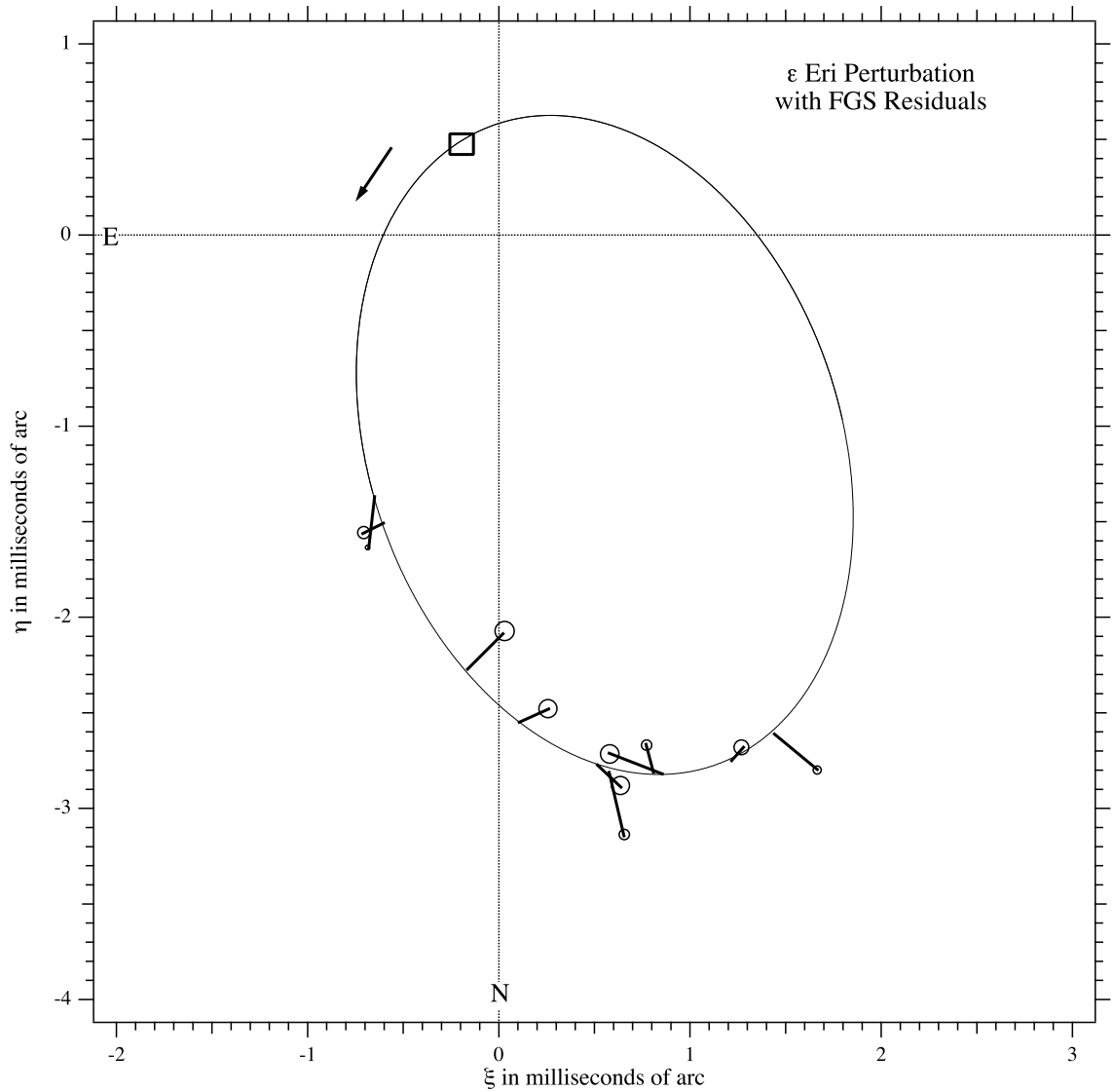}
\caption{Perturbation orbit for \eps. Elements are found in Table~\ref{tbl-ORB}.  
Residual vectors are plotted, connecting each normal point residual to its predicted
position at each epoch of observation. Circle size is proportional to the number 
of observations forming the normal point. The square marks periastron passage, 
T$_0$=2007.3. The arrow indicates the direction of motion of \eps.} \label{fig-OrbResids}

\end{figure}

We find a perturbation size, $\alpha_A = 1.88 \pm 0.19$ mas, and an inclination, 
$i $= 30\fdg1 $\pm$ 3\fdg8. These, and the other orbital elements for the 
perturbation, are listed in Table~\ref{tbl-ORB} with 1-$\sigma$ errors. 
Errors generated by GaussFit (Jefferys \etal 1988) come from
a maximum likelihood estimation that is an 
approximation to a Bayesian maximum a posteriori 
estimator with a flat prior (Jefferys 1990).  Figure~\ref{fig-PJconsTEMP} illustrates the Pourbaix and Jorrisen relation (Equation 10) between parameters 
obtained from astrometry (left-side) and radial velocities (right side) and our final  
estimates for $\alpha_A$ and $i$. As seen in Tables~\ref{tbl-SUM} and \ref{tbl-ORB}, most of the errors of the terms in Equation 10 are quite small. In essence, our simultaneous solution uses the 
Figure~\ref{fig-PJconsTEMP} curve as a quasi-Bayesian prior, sliding along it until the 
astrometric and radial velocity residuals are minimized. Gross deviations from the curve are 
minimized by the high precision of many of the terms in Equation 10.   Figure~\ref{fig-RVS} contains 
all radial velocity measures and the predicted velocity curve from the simultaneous 
solution. Compared to the typical perturbation radial velocity curve (e.g. Hatzes \etal 2005, McArthur \etal 2004, Cochran \etal 2004), Figure~\ref{fig-RVS} exhibits far more scatter about the derived orbit. There are two reasons for this. The perturbation amplitude is small (K1=18.5 \ms), and \eps~ is an active star, as discussed in Hatzes \etal (2000). Reiterating their conclusions, none of the activity cycles have periods commensurate with the planetary perturbation period. Figure~\ref{fig-OrbResids}  presents the astrometric residuals and the derived 
perturbation orbit for the primary star, \eps. Stellar activity has even less of an effect on astrometry at our level of precision. A star spot covering 30\%  of the surface would induce a photocenter shift of less than 0.2 mas (Sozzetti 2005). The astrometry confirms the existence of the companion.

Our analysis of the radial velocities (now spanning over 25 years, all shown in Figure~\ref{fig-RVS}) included a linear drift term, a change in velocity as a function of time. This drift is clearly seen in the overplotted final radial velocity curve, and amounts to 0.32 $\pm$ 0.05 \ms yr$^{-1}$. Such a change can be caused by longer-period companions and/or secular acceleration \citep{Kur04}. The secular acceleration expected for \eps~is (van de Kamp, 1967) quite small, 0.07 \ms yr$^{-1}$. The trend we find is over four times larger than the predicted secular acceleration,  and is approximately the acceleration one might find for a planet similar in mass to \eps~ b, but with a 50--100 year period.   Figure 9 shows sufficient enough overlap among 
the many velocity data sets that the trend is unlikely
an artifact due to a mismatch in the center of mass velocity offsets
(discussed above) obtained for each set. Typical offset random error is $\sim1$ \ms.
While this acceleration is not a detection of the longer-period companion ($40< a <60$ AU) invoked by \nocite{QT02,Oze00} Quillan \& Thorndike (2002) and Ozernoy \etal (2000) to modify the dust distribution as discussed below in Section 7, it may (with a semimajor axis 10--20 AU) be at least partially responsible for the inner cavity in the dust disk distribution imaged by Greaves \etal (2005). The astrometric motion  over 15 years due to this possible tertiary would be of order 3 mas and difficult to separate from proper motion, e.g., \citet{Bla82}.

The planetary mass depends on the mass of the primary star, for which we have 
adopted $M_* = 0.83\pm0.05M_{\sun}$ \citep{DiF04}. For this $M_*$ we find $M_b = 
1.55\pm0.24M_{Jup}$. The companion is clearly an extrasolar giant planet. In 
Table~\ref{tbl-massprm} the mass value, $M_b$, incorporates the present uncertainty 
in  $M_*$.  Until \eps~b is directly detected, its radius is unknown. From a review of exoplanet masses and radii \citep{Gui05}, a radius of $R = 1 R_{Jup}$ seems reasonable.

Our eccentricity value, $e$ = 
0.70, allows for a significant  difference in separation between star and exoplanet at 
apastron compared to  periastron. At time of periastron passage, T$_0$ = 2007.29, we 
predict a separation 0\farcs3 $\pm$ 0\farcs1 at a position angle of -27\arcdeg. At the 
next apastron, to occur 2010.71, the separation should be 1\farcs8  $\pm$ 0\farcs4 at 
position angle of 153\arcdeg. The dominant sources of error for the separations are the 
eccentricity (6\%)  and the \eps~b planet mass (15\%).

\section{Discussion}

Our accurate mass 
and orbital parameters for this planetary companion have value for future direct 
imaging  projects. We now know where near \eps~to look for \eps~b. We would now like to know when to look, what 
bandpass is best, and what we can expect to see. As stated previously system age, companion mass, and orbital geometry are critical parameters when estimating visibility. 

A high level of chromospheric activity is seen for \eps~(e.g. Gray \& Baliunas 1995), and  is consistent with a relatively young age; $<$ 1 Gyr (Soderblom \& D\"appen 1989). 
 Saffe \etal (2005) used the calibrations of Donahue (1993) and Rocha-Pinto \& Maciel (1998)  (which corrected the age with an effect from stellar metallicity) to estimate ages of 0.66 and  0.82 Gyr, respectively.  Henry (1986) derived from CaII lines a value of 0.8 Gyr. Song \etal (2000) used Li abundances with its position in the H-R diagram and kinematics to derive a value of 0.73 $\pm$ 0.2 Gyr.
 Di Folco \etal (2004) \nocite{DiF04} estimated the age at  0.85 Gyr, a value obtained through the measurement of the radius of \eps~by long-baseline interferometry. Their modeling is consistent with a  primary mass of $M_* = 0.83\pm0.05M_{\sun}$, an estimate that weakly depends on measured metallicity, which ranges -0.13 $<$ [Fe/H] $<$ -0.06 in the literature.
 
Hubbard \etal (2002) predict the intrinsic luminosity of extrasolar giant 
planets as a function of mass and age. From their Figure 11, an age of 800 My \cite{DiF04}, 
and our planetary mass, $M_b = 1.55\pm0.24M_{Jup}$, 
we find for \eps~b, $L/L_{\sun} = 1.6\times10^{-8}$.Using the Di Folco \etal (2004) T$_{eff}$=5135 K, their radius, $R_* = 0.743R_{\sun}$, our parallax, $\pi_{abs} = 311.37$ mas, and a bolometric correction, B.C. = -0.27, from Flower (1996)\nocite{Flo96}, we find a difference in bolometric magnitude of \eps~compared to the Sun of $\Delta M_{bol} = +1.17$. Hence, neglecting reflected light 
and orbital phase, \eps~b is 4.67$\times10^{-8}$ fainter in bolometric luminosity than 
\eps.

Sudarsky \etal (2005), Dyudina \etal 
(2005), and Burrows \etal (2004) discuss exoplanet apparent brightness in reflected host 
star light as functions of orbit geometry, orbital phase, and cloud cover. Burrows \etal 
(2004) predict the full spectrum of \eps~b  from 0.5 to 6$\micron$, asserting that the planet 
is too young for its atmosphere to contain condensed ammonia clouds. However, \eps~b should 
exhibit H$_2$O clouds. They predict a maximum planet/host star flux ratio, 
log$_{10}(F_{planet}/F_{star}) \sim-7$, at $\sim4.5\micron$ with a secondary 
peak, log$_{10}(F_{planet}/F_{star}) \sim-8$, at $\sim1\micron$. Dyudina \etal (2005) 
predict that for $\omega = 30\arcdeg$, inclination,  $i = 30\arcdeg$, $e$ = 0.5, 
and a Jupiter-like atmosphere, the planet/host star flux ratio is largest very shortly 
after periastron, late 2007. However, the separation remains small ($\sim$0\farcs3). The inclination of the \eps~system, $i = 30\arcdeg$,  is 
likely to decrease the flux ratio by approximately a factor of two (Sudarsky \etal 2005), 
compared to a $i = 90\arcdeg$ edge-on orientation. Given the orientation of the orbit of  
\eps~b (its ascending node, $\Omega$' = 254\arcdeg), the disk of \eps~b is most fully 
illuminated at apastron, but is three times further away from its primary.

The dusty rings or debris disks surrounding \eps~also suggest relative youth for the system. Photometric measurements from the IRAS satellite (Aumann, 1988) provided the first hint of dust 
around \eps.  Subsequently, Submillimeter Common-User Bolometric Array (SCUBA) measurements were made between 1997 and 2002.    These measurements determined that the dust, distributed in a ring, is  located 65 AU from the star \citep{Gre98,Gre05}.   The sub-mm bolometer, SIMBA, provided observational confirmation of this extended dust disk (Schutz \etal 2004).    The STIS CCD camera on \HST took deep optical images
around \eps~in an effort to find an optical counterpart for the sub-millimeter observations.  These 
measurements did not provide clear evidence for the detection of  that optical counterpart, but did
place a limit on  the optical surface brightness of the dust, that it could not be brighter than approximately 25 STMAG arcsec$^{-2}$, which places constraints on the nature and amount of the smallest dust grains (Proffit \etal 2004).

Observational and theoretical searches for the signature of planetary/brown dwarf objects in the structure of the dust disk around \eps~are underway.  Clumps seen in 
the ring are thought to come from the interaction between the disk and a massive
planetary body (Holland \etal 2003).  Adaptive optics on the
Keck Telescope were used to search for extrasolar planets.  These studies found no 
evidence of brown dwarf or planetary companions down to 5 Jupiter masses at the 
angular separations comparable to that of the dust rings (Macintosh \etal 
2003).   Spitzer Space Telescope (SST) observations  made with the Multi-band Imaging Photometer (MIPS) and the InfrRed Spectograph (IRS)  have confirmed the disk and provided evidence for asymmetries in the structure of the disk that may have been caused by the gravitational perturbation of sub-stellar companions (Marengo \etal 2004).  

Two recent studies suggested that debris disks and long-period planets co-exist, with planetary bodies 'sculpting' the disk. \eps~is the prototypical system.  First, high-resolution modeling of the structure of the disk around \eps~predict an angular motion of the asymmetry of the disk of about 0\fdg6-0\fdg8 yr$^{-1}$ (Ozervnoy \etal 2000). Secondly,  Quillen \& Thorndike (2002) carried out numerical simulations of dust particles captured in mean motions resonances with a
hypothetical planet (e = 0.3 , M = $10^{-4}$ \msun, a = 40 AU) at periastron. These produced a dust  distribution that agreed with the morphology of the dust ring around \eps~presented by Greaves \etal (1998, 2005). An investigation into the dynamics of the dust ring around \eps~(Moran, Kuchner \& Holman, 2004) concluded that the eccentricity of the dust released in the inner ring ($<$20 AU)  
could reveal patterns in the dust which could confirm the existence of the planet reported by Hatzes \etal (2000).  
%Recent sub-millimeter observations suggest perturbations by a planet orbiting at tens of AU, and note that the center is relatively clear of dust (Greaves \etal 2005). This latter feature could be a corroborating signature of the tertiary we deduce from the observed linear trend in the radial velocity.

We determined an inclination of $i$=30\fdg1 $\pm$ 3\fdg2  for \eps~b.   Our measured inclination is consistent with the previously measured dust disk inclination from 450 and 850 $\mu$m maps of \cite{Gre98,Gre05}, $i = 25$\arcdeg. This suggests that the dust disk 
and plane of the orbit of \eps~b are coincident and that the dust distribution is nearly 
circular. This provides support for hiearchical accretion models for planet formation 
\cite{Pol96}, where coplanar dust and a debris disk are expected remnants of 
planet formation (Tsiganis \etal 2005).   Lastly, \eps~b and the possible tertiary deduced from the linear trend in the radial velocities (Section 6) would most likely eject particles that would spiral inward, and recent SCUBA submillimeter observations have  shown that the center of the disk is relatively excavated of dust, with half or less of the signals seen in the ring (Greaves \etal 2005).

\section{Conclusions}

	Analyzing three years of {\it HST} FGS and over 14 years of Allegheny 
Observatory MAP astrometry, we find an  independently determined parallax and proper motion for  \eps~ that agree within the errors with HIPPARCOS.

	Astrometric observations with \HST FGS, combined with long-duration Allegheny  MAP astrometry and ground based radial velocities, have  confirmed the existence of the planet  orbiting \eps, first suggested by Walker \etal (1995), noted by Cummings \etal (1999), and finally announced by Hatzes \etal(2000). 

	Combining the astrometry with radial velocities from six different sources, 
spanning 25 years, and applying the Pourbaix \& Jorrisen constraint between astrometry 
and radial velocities, we obtain for the perturbing object \eps~b a period, P=6.85 $\pm$ 
0.02 y, inclination, $i$=30\fdg1 $\pm$ 3\fdg2, and perturbation semimajor axis, 
$\alpha_A = 1.88 \pm 0.19$ mas.  Assuming for \eps~a stellar mass $M_*$ = 0.83$\pm$0.05$M_{\sun}$, we obtain a mass for \eps~b, $M_b$ = 1.55$\pm$0.24$M_{Jup}$.  This companion inclination  matches the disk inclination determined by Greaves \etal(2005).

Our astrometry predicts for \eps~b periastron passage at T$_0$ = 2007.29, 
with a separation $\sim$0\farcs3 in position angle -27\arcdeg (a = 3.39 AU). At the next apastron, to 
occur 2010.71, the separation should be 1\farcs7 in position angle 153\arcdeg. The 
orbital geometry suggests that 2007.97 (late December 2007) is the most favorable time 
for direct detection in reflected light.   For an \eps~ age $\sim$850 My and our determined mass, \eps~b will have an intrinsic luminosity  $L/L_{\sun} = 1.6\times10^{-8}$\nocite{Hub02}, 4.67$\times10^{-8}$ 
times fainter in bolometric luminosity than \eps.

Radial velocities spanning 25 years indicate a long-term linear trend, an acceleration consistent with a Jupiter-mass object with a period of 50--100 years. This is a possible detection of a tertiary companion responsible for a major feature of the dust morphology, the central cavity.

\acknowledgments

Support for this work was provided by NASA through grants 
GO-09167, GO-09347, and GO-09969 from the Space Telescope 
Science Institute, which is operated
by the Association of Universities for Research in Astronomy, Inc., under
NASA contract NAS5-26555.  The Allegheny Observatory received support from the National Science Foundation through grant AST-0098552 and the National Aeronautics and Space Administration through grant NAG-510628.
 This publication makes use of data products from the 
Two Micron All Sky Survey, which is a joint project of the University of 
Massachusetts 
and the Infrared Processing and Analysis Center/California Institute of 
Technology, 
funded by NASA and the NSF.  This research has made use of the SIMBAD database, 
operated at CDS, Strasbourg, France; the NASA/IPAC Extragalactic Database (NED) 
which is operated by JPL, California Institute of Technology, under contract 
with the 
NASA;  and NASA's Astrophysics Data System Abstract Service.  We thank an anonymous referee for a careful review and constructive criticism which improved this final version of the paper.

\clearpage

% Now comes the reference list.  

%\clearpage

\clearpage

\begin{center}
%\begin{tiny}
\begin{deluxetable}{rlllrlll}
\tablewidth{5in}
\tablecaption{Log of \eps ~ FGS Observations\label{tbl-LOOF}}
\tablehead{\colhead{Epoch}&
\colhead{MJD\tablenotemark{a}}&\colhead{Year}&
\colhead{Roll (\arcdeg)\tablenotemark{b}}&
\colhead{Epoch}&
\colhead{MJD}&\colhead{Year}&
\colhead{Roll (\arcdeg)}}
\startdata
1&51946.528&2001.101&96.9&24&52862.76878&2003.609&281.0\\
2&51982.209&2001.198&91.0&25&52865.68038&2003.617&280.0\\
3&52180.393&2001.741&280.0&26&52868.70565&2003.625&280.0\\
4&52309.062&2002.093&97.0&27&52871.75072&2003.634&280.0\\
5&52497.787&2002.610&280.0&28&52874.84206&2003.642&280.0\\
6&52500.124&2002.616&280.0&29&53007.24013&2004.005&106.0\\
7&52503.062&2002.624&280.0&30&53010.26367&2004.013&105.0\\
8&52506.667&2002.634&280.0&31&53014.10664&2004.024&105.0\\
9&52509.629&2002.642&280.0&32&52672.409&2003.088&105.0\\
10&52513.545&2002.653&280.0&33&52679.324&2003.107&105.0\\
11&52537.069&2002.718&280.0&34&52681.993&2003.114&105.0\\
12&52540.651&2002.727&280.0&35&52682.004&2003.114&105.0\\
13&52543.679&2002.736&280.0&36&52683.127&2003.117&105.0\\
14&52642.962&2003.007&105.0&37&52686.352&2003.126&105.0\\
15&52645.364&2003.014&105.0&38&52688.330&2003.132&105.0\\
16&52649.703&2003.026&105.0&39&52862.769&2003.609&281.0\\
17&52672.40928&2003.088&105.0&40&52865.680&2003.617&280.0\\
18&52679.32437&2003.107&105.0&41&52868.706&2003.625&280.0\\
19&52681.9927&2003.114&105.0&42&52871.751&2003.634&280.0\\
20&52682.00362&2003.114&105.0&43&52874.842&2003.642&280.0\\
21&52683.12669&2003.117&105.0&44&53007.240&2004.005&106.0\\
22&52686.35236&2003.126&105.0&45&53010.264&2004.013&105.0\\
23&52688.32956&2003.132&105.0&46&53014.107&2004.024&105.0\\
\enddata
\tablenotetext{a}{MJD = JD - 2400000.5}\\
\tablenotetext{b}{Spacecraft roll as defined in Chapter 2, FGS Instrument 
Handbook \cite{Nel03}}
\end{deluxetable}
%\end{tiny}
\end{center}

\begin{center}
\begin{deluxetable}{rllrll}
\tablewidth{5in}
\tablecaption{Log of \eps ~MAP Observations\label{tbl-LOOM}}
\tablehead{\colhead{Epoch}&
\colhead{MJD\tablenotemark{a}}&
\colhead{Year}&
\colhead{Epoch}&
\colhead{MJD}&
\colhead{Year}}
\startdata
1&47544.0781&1989.0474&66&53001.0760&2003.988\\
2&47772.4045&1989.6726&67&50362.2622&1996.763\\
3&47790.3559&1989.7217&68&50362.2948&1996.763\\
4&47822.2920&1989.8091&69&50362.3281&1996.763\\
5&47826.2622&1989.8200&70&50371.2538&1996.788\\
6&47829.2503&1989.8282&71&50371.2872&1996.788\\
7&47905.0385&1990.0357&72&50371.3184&1996.788\\
8&47923.0024&1990.0849&73&50719.3260&1997.741\\
9&47933.0142&1990.1123&74&50719.3594&1997.741\\
10&48209.2163&1990.8685&75&50736.3129&1997.787\\
11&48213.2052&1990.8794&76&50736.3448&1997.787\\
12&48234.1406&1990.9367&77&50741.3281&1997.801\\
13&48281.0559&1991.0652&78&50798.1108&1997.956\\
14&48290.0135&1991.0897&79&50799.1316&1997.959\\
15&48528.3413&1991.7422&80&50799.1649&1997.959\\
16&48547.2948&1991.7941&81&51079.3615&1998.727\\
17&48901.3066&1992.7633&82&51079.3934&1998.727\\
18&48915.2816&1992.8016&83&51079.4253&1998.727\\
19&49007.9990&1993.0554&84&51100.3219&1998.784\\
20&49022.0066&1993.0938&85&51100.3538&1998.784\\
21&49027.0066&1993.1075&86&51145.1372&1998.907\\
22&49268.2733&1993.7680&87&51145.1698&1998.907\\
23&49274.2691&1993.7844&88&51145.2024&1998.907\\
24&49285.2309&1993.8145&89&51209.0010&1999.081\\
25&49302.2240&1993.8610&90&51829.2788&2000.780\\
26&49312.1545&1993.8882&91&51829.3115&2000.780\\
27&49334.1448&1993.9484&92&51829.3434&2000.780\\
28&49372.9997&1994.0548&93&52183.3115&2001.749\\
29&49597.4087&1994.6692&94&52183.3448&2001.749\\
30&49600.4003&1994.6773&95&52185.2774&2001.754\\
31&49640.3052&1994.7866&96&52219.1969&2001.847\\
32&49653.2205&1994.8220&97&52219.2316&2001.847\\
33&49668.1733&1994.8629&98&52219.2628&2001.847\\
34&49708.0587&1994.9721&99&52219.3080&2001.848\\
35&49747.0052&1995.0787&100&52219.3413&2001.8476\\
36&49748.0476&1995.0816&101&52220.2115&2001.8500\\
37&50362.2622&1996.7632&102&52220.2455&2001.8501\\
38&50371.2538&1996.7878&103&52225.1816&2001.8636\\
39&50719.3260&1997.7408&104&52225.2149&2001.8637\\
40&50736.3129&1997.7873&105&52225.2483&2001.8638\\
41&50741.3281&1997.8010&106&52226.1899&2001.8664\\
42&50798.1108&1997.9565&107&52226.2233&2001.8665\\
43&50799.1316&1997.9593&108&52226.2573&2001.8665\\
44&51079.3615&1998.7265&109&52227.2226&2001.8692\\
45&51100.3219&1998.7839&110&52227.2552&2001.8693\\
46&51145.1372&1998.9066&111&52265.0691&2001.9728\\
47&51209.0010&1999.0815&112&52265.1024&2001.9729\\
48&51829.2788&2000.7797&113&52265.1351&2001.9730\\
49&52183.3115&2001.7490&114&52893.3497&2003.6929\\
50&52185.2774&2001.7544&115&52893.383&2003.6930\\
51&52219.1969&2001.8472&116&52924.2726&2003.7776\\
52&52219.3080&2001.8475&117&52924.3059&2003.7777\\
53&52220.2115&2001.8500&118&52925.2747&2003.7804\\
54&52225.1816&2001.8636&119&52925.3094&2003.7805\\
55&52226.1899&2001.8664&120&52925.342&2003.7805\\
56&52226.2573&2001.8665&121&52925.3774&2003.7806\\
57&52227.2226&2001.8692&122&52937.2087&2003.8130\\
58&52265.0691&2001.9728&123&52937.2413&2003.8131\\
59&52893.3497&2003.6929&124&52946.224&2003.8377\\
60&52924.2726&2003.7776&125&52946.2566&2003.8378\\
61&52925.2747&2003.7804&126&52946.2899&2003.8379\\
62&52925.3420&2003.7805&127&53000.058&2003.9851\\
63&52937.2087&2003.8130&128&53001.076&2003.9879\\
64&52946.2240&2003.8377&129&53001.1177&2003.9880\\
65&53000.0580&2003.9851&130&53001.1448&2003.9881\\
\enddata
\tablenotetext{a}{MJD = JD - 2400000.5}\\
\end{deluxetable}
\end{center}
\clearpage

\begin{deluxetable}{lllcc}
\tablewidth{4in}
\tablecaption{The Radial Velocity Data Sets \label{tbl-RV}}
\tablehead{ 
\colhead{Data Set}& 
\colhead{Coverage}& 
\colhead{Technique} & 
\colhead{N} & 
\colhead{RMS} \\& 
\colhead{(yr)}&  &  & 
\colhead{(\ms)} }
\startdata
CFHT  & 1980.81-1991.88 & HF cell &48   & 10.5 \\
Lick  & 1987.69-1998.99 &Iodine cell& 54   & 11.5  \\
McD $\phi$I  &1988.74-1994.81 & Telluric & 27   & 15.2  \\
McD $\phi$II  &1990.78-1998.07 &Iodine cell &42   &  11.7  \\
ESO  & 1992.84-1998.02 &Iodine cell &  36   & 9.6  \\
McD $\phi$III & 1998.69-2004.86 & Iodine cell & 28 &7.4  \\
\hline
&&total&235&
\enddata
\end{deluxetable}

%\clearpage
\begin{deluxetable}{lrlr}
\tablewidth{4in}
\tablecaption{New McDonald $\phi$III Radial Velocities \label{tbl-P3}}
\tablehead{ 
\colhead{mJD}& 
\colhead{RV$_{\phi III}$}& 
\colhead{mJD} & 
\colhead{RV$_{\phi III}$}  \\
& 
\colhead{(\ms)}&  &
\colhead{(\ms)} }
\startdata
51066.4339&-3.6$\pm$6.4&52539.4226&-16.4$\pm$6.4\\
51212.1671&13.6 5.0&52576.4378&2.0 5.5\\
51239.1133&1.7 8.6&52661.1445&-11.8 5.2\\
51449.4333&6.5 5.0&52931.3912&-13.9 6.4\\
51503.3574&14.7 5.1&52958.2449&-0.4 5.4\\
51529.1986&-3.5 7.4&52958.2481&0.0 6.0\\
51555.1645&-0.2 5.2&53016.2424&2.9 6.0\\
51775.4643&1.3 5.9&53016.2456&2.1 6.0\\
51809.4033&-2.7 4.8&53035.1757&-5.2 5.6\\
51917.2260&2.9 5.5&53075.0940&3.0 5.7\\
51984.0740&-2.8 5.0&53318.3124&-10.1 6.6\\
52142.4227&-0.7 6.2&53632.4520&-1.2 5.3\\
52142.4264&-0.4 6.1&53632.4550&0.4 5.5\\
52248.2921&-14.0 5.9&53689.3900&-3.0 5.2\\
52303.1266&-16.1 7.7&53745.2110&6.3 5.1\\
52328.1208&2.1 5.6&53809.0850&9.0 5.5\\
52330.1064&-0.9 5.7&&\\

\enddata
\end{deluxetable}
\clearpage

\begin{deluxetable}{rlllll}
\tablewidth{0in}
\tablecaption{FGS1r and MAP Astrometric Reference Stars\label{tbl-2MASSPos}}
\tablehead{
\colhead{ID}&
\colhead{Catalog}
&\colhead{RA\tablenotemark{a}~~~ (2000.0)~}&
\colhead{Dec\tablenotemark{a}}&
\colhead{V\tablenotemark{b}}&\colhead{2MASS}
}
\startdata
1&\eps&53.232961&-9.458295&3.82&03325591-0927298\\
2&&53.325845&-9.467569&15.61&03331820-0928032\\
3&&53.312638&-9.434314&15.58&03331503-0926035\\
4&&53.295306&-9.421752&16.12&03331087-0925183\\
5&&53.269681&-9.418811&16.41&03330472-0925077\\
6&&53.207917&-9.449667&16.48&\\
\hline
7&GEN\# +6.10280864&53.374062&-9.504524&11.36&03332977-0930162\\
8&HD 22130&53.401866&-9.349848&9.52&03333644-0920594\\
9&BD-09 696 &53.164328&-9.267778&10.69&03323943-0916040\\
10&GEN\# +6.10280861 &53.032108&-9.514245&11.60&03320770-0930512\\
11&BD-10 699 &53.399013&-9.586081&10.24&03333576-0935098\\
12&HD 21951  &53.001248&-9.385150&9.69&03320029-0923065\\
13&BD-10 695 &53.320908&-9.696887&9.78&03331701-0941487\\
14&BD-10 700 &53.466004&-9.641330&9.94&03335184-0938287\\
15&BD-09 699  &53.429559&-9.232722&11.20&03334309-0913577\\
16&2MASS 03320556-0945292&53.023188&-9.758118&11.15&03320556-0945292\\

\enddata
\tablenotetext{a}{Positions from 2MASS, except ID 6 from applying STScI Visual Target Tuner to the Digital Sky Survey.}
\tablenotetext{b}{Magnitudes from FGS1r (ID 2-6), SIMBAD (ID 1), or MAP (ID 7-
16).}
\end{deluxetable}

%TABLE 4
\begin{deluxetable}{cccccc}
\tablewidth{0in}
\tablecaption{V and Near-IR Photometry \label{tbl-IR}}
\tablehead{\colhead{ID}&
\colhead{$V$} &
\colhead{$K$} &
\colhead{$(J-H)$} &
\colhead{$(J-K)$} &
\colhead{$(V-K)$} 
}
\startdata
1&3.82$\pm$0.01&1.82$\pm$0.05&0.40$\pm$0.28&0.48$\pm$0.07&2.00$\pm$0.05\\
2&15.61$\pm$0.03&13.01$\pm$0.03&0.58$\pm$0.04&0.72$\pm$0.04&2.60$\pm$0.04\\
3&15.58$\pm$0.03&14.09$\pm$0.06&0.41$\pm$0.05&0.44$\pm$0.07&1.50$\pm$0.06\\
4&16.12$\pm$0.03&13.88$\pm$0.05&0.51$\pm$0.04&0.59$\pm$0.06&2.24$\pm$0.06\\
5&16.41$\pm$0.03&14.21$\pm$0.08&0.46$\pm$0.07&0.62$\pm$0.09&2.20$\pm$0.08\\
6&16.48&&&&\\
\hline
7&11.36$\pm$0.03&9.93$\pm$0.02&0.29$\pm$0.03&0.33$\pm$0.03&1.43$\pm$0.04\\
8&9.52$\pm$0.01&8.63$\pm$0.02&0.21$\pm$0.06&0.24$\pm$0.03&0.89$\pm$0.02\\
9&10.69$\pm$0.03&8.85$\pm$0.02&0.48$\pm$0.04&0.54$\pm$0.03&1.84$\pm$0.04\\
10&11.60$\pm$0.03&10.43$\pm$0.02&0.33$\pm$0.03&0.41$\pm$0.03&1.17$\pm$0.0
4\\
11&10.24$\pm$0.03&8.73$\pm$0.02&0.40$\pm$0.06&0.43$\pm$0.04&1.51$\pm$0.04\\
12&9.69$\pm$0.03&8.98$\pm$0.02&0.13$\pm$0.03&0.18$\pm$0.03&0.71$\pm$0.04\\
13&9.78$\pm$0.01&7.87$\pm$0.02&0.47$\pm$0.03&0.55$\pm$0.02&1.91$\pm$0.02\\
14&9.94$\pm$0.03&7.69$\pm$0.03&0.57$\pm$0.06&0.66$\pm$0.04&2.25$\pm$0.04\\
15&11.20$\pm$0.03&8.13$\pm$0.03&0.62$\pm$0.05&0.74$\pm$0.03&3.07$\pm$0.04\\
16&11.15$\pm$0.03&9.43$\pm$0.02&0.45$\pm$0.03&0.49$\pm$0.03&1.72$\pm$0.04\\
\enddata
\end{deluxetable}

%TABLE 6
\begin{deluxetable}{lccccc}
\tablewidth{0in}
\tablecaption{Astrometric Reference Star Adopted
Spectrophotometric Parallaxes \label{tbl-SPP}}
\tablehead{\colhead{ID}& \colhead{Sp. T.\tablenotemark{a}}&
\colhead{V} & \colhead{M$_V$} &\colhead{m-M}& 
\colhead{$\pi_{abs}$(mas)}} 
\startdata
2&K4V&15.6&7.1&8.5$\pm$1&1.9$\pm$1.0\\
3&G8V&15.6&5.6&10.0$\pm$0.4&1.0$\pm$0.2\\
4&K2V&16.1&6.5&9.6$\pm$0.4&1.2$\pm$0.2\\
5&K2V&16.4&6.5&9.9$\pm$0.4&1.0$\pm$0.2\\
6&K2V&16.4&6.5&9.9$\pm$2&1.0$\pm$0.9\\
\hline
7&G0V&11.4&4.4&7.0$\pm$0.4&4.1$\pm$0.7\\
8&F5V&9.5&3.5&6.0$\pm$0.4&6.3$\pm$1.2\\
9&K0V&10.7&5.9&4.8$\pm$0.4&11.0$\pm$2.0\\
10&G5V&11.6&5.1&6.5$\pm$0.4&5.0$\pm$0.9\\
11&G8V&10.2&5.6&4.6$\pm$0.4&11.8$\pm$2.2\\
12&F0V&9.7&2.7&7.0$\pm$0.4&4.0$\pm$0.7\\
13&K0V&9.8&5.9&3.9$\pm$0.4&16.7$\pm$3.1\\
14&K0III&9.9&0.7&9.2$\pm$0.4&1.4$\pm$0.3\\
15&K2III&11.2&2.7&8.5$\pm$0.4&2.0$\pm$0.4\\
16&G8V&11.2&5.6&5.6$\pm$0.4&7.8$\pm$1.4\\

\enddata 
\tablenotetext{a}{Spectral types and luminosity class estimated from colors and 
reduced 
proper motion diagram.}\\
\end{deluxetable}

%TABLE 7
\begin{deluxetable}{cccc}
\tablewidth{0in}
\tablecaption{\eps ~ and Reference Star Relative Positions    \label{tbl-POS}}
\tablehead{\colhead{ID}&
\colhead{$V$} &
\colhead{$\xi$\tablenotemark{a}} &
\colhead{$\eta$\tablenotemark{a}} 
}
\startdata
1\tablenotemark{b}&3.73&297.0383$\pm$0.0001&35.6893$\pm$0.0001\\
2&15.57&-25.6479$\pm$0.0002&126.1290$\pm$0.0001\\
3\tablenotemark{c}&15.57&0.0000$\pm$0.0002&0.0000$\pm$0.0002\\
4&16.09&52.8710$\pm$0.0003&-55.1822$\pm$0.0002\\
5&16.40&433.1267$\pm$0.0003&-20.0260$\pm$0.0003\\
6&16.37&140.6876$\pm$0.0004&-81.2800$\pm$0.0002\\
\hline
1\tablenotemark{d}&3.73&715.2729$\pm$0.0002&201.7543$\pm$0.0002\\
7&11.36&1214.0919$\pm$0.0013&35.0491$\pm$0.0014\\
8&9.52&1313.1141$\pm$0.0005&591.8986$\pm$0.0007\\
9&10.69&469.2508$\pm$0.0011&887.5650$\pm$0.0014\\
10\tablenotemark{e}&11.60&0.0000$\pm$0.0021&0.0000$\pm$0.0025\\
11&10.24&1302.5205$\pm$0.0007&-258.4640$\pm$0.0011\\
12&9.69&-109.9029$\pm$0.0008&464.7282$\pm$0.0008\\
13&9.78&1025.1161$\pm$0.0005&-657.0894$\pm$0.0007\\
14&9.94&1540.2424$\pm$0.0005&-457.4732$\pm$0.0006\\
15&11.2&1411.6967$\pm$0.0011&1013.5480$\pm$0.0018\\
16&11.15&-31.1988$\pm$0.0015&-877.8886$\pm$0.0023\\

\enddata
\tablenotetext{a}{$\xi$ and $\eta$ are relative positions in arcseconds
}
\tablenotetext{b}{epoch 2002.614, J2000}
\tablenotetext{c}{RA = 53.312638, Dec = -9.434314, J2000}
\tablenotetext{d}{epoch 1996.761, J2000}
\tablenotetext{e}{RA = 53.032108, Dec = -9.514245, J2000}

\end{deluxetable}

%TABLE 8
\begin{deluxetable}{cccccc}
\tablewidth{0in}
\tablecaption{Reference Star Proper Motions \label{tbl-PM}}
\tablehead{\colhead{}&
\colhead{}&
\colhead{Input (UCAC2)} &
\colhead{}&
\colhead{Final ({\it HST})} &
\colhead{}  \\
\colhead{ID}&
\colhead{V} &
\colhead{$\mu_\alpha$\tablenotemark{a}} &
\colhead{$\mu_\delta$\tablenotemark{a}}&
\colhead{$\mu_\alpha$\tablenotemark{a}} &
\colhead{$\mu_\delta$\tablenotemark{a}} }
\startdata 
2&15.57&0.0053$\pm$0.0078&-0.0112$\pm$0.008&0.0110$\pm$0.0002&-
0.0076$\pm$0.0002\\
3&15.57&0.0166$\pm$0.0076&-0.0014$\pm$0.0076&0.0072$\pm$0.0003&-
0.0026$\pm$0.0002\\
4&16.09&0.0044$\pm$0.009&-0.0028$\pm$0.0079&0.0080$\pm$0.0004&-
0.0022$\pm$0.0002\\
5&16.4&-0.0004$\pm$0.004&-0.0044$\pm$0.004&0.0070$\pm$0.0004&
0.0082$\pm$0.0002\\
6&16.37&0.0064$\pm$0.003&0.0083$\pm$0.003&0.0014$\pm$0.0004&-0.0030$\pm$0.
0003\\
\hline
7&11.36&0.0253$\pm$0.005&0.0117$\pm$0.005&0.0221$\pm$0.0002&0.0083$\pm$0.
0003\\
8&9.52&0.0104$\pm$0.005&-0.0025$\pm$0.005&0.0088$\pm$0.0001&-
0.0052$\pm$0.0001\\
9&10.69&0.003$\pm$0.005&-0.0178$\pm$0.005&0.0035$\pm$0.0002&-
0.0184$\pm$0.0003\\
10&11.6&-0.0015$\pm$0.005&-0.0074$\pm$0.005&-0.0020$\pm$0.0004&-
0.0029$\pm$0.0004\\
11&10.24&0.0101$\pm$0.005&-0.0094$\pm$0.005&0.0126$\pm$0.0001&-
0.0089$\pm$0.0002\\
12&9.69&0.0095$\pm$0.005&-0.0038$\pm$0.005&0.0109$\pm$0.0001&-
0.0074$\pm$0.0002\\
13&9.78&0.0549$\pm$0.005&-0.037$\pm$0.005&0.0579$\pm$0.0001&-
0.0326$\pm$0.0001\\
14&9.94&-0.0003$\pm$0.005&-0.0079$\pm$0.005&-0.0017$\pm$0.0001&-
0.0102$\pm$0.0001\\
15&11.2&0.0117$\pm$0.005&-0.0063$\pm$0.005&0.0108$\pm$0.0002&-
0.0042$\pm$0.0003\\
16&11.15&0.0073$\pm$0.005&-0.0136$\pm$0.005&0.0052$\pm$0.0003&-
0.0158$\pm$0.0004\\
\enddata

\tablenotetext{a}{$\mu_\alpha$ and $\mu_\delta$ are relative motions in arcsec
yr$^{-1}$ }
\end{deluxetable}

%TABLE 9
%\begin{center}
\begin{deluxetable}{rlll}
\tablewidth{0in}
\tablecaption{\eps~ Parallax and Proper Motion \label{tbl-SUM}}
\tablewidth{0in}
\tablehead{\colhead{Parameter} &  \colhead{\HST} &  \colhead{MAP} &  
\colhead{Combined}}
\startdata
Study duration  &2.92 y & 14.94 y& \\
number of observation sets    &   46  & 130 &\\
reference star $\langle V\rangle$ &  16.0 & 10.2 & 12.2    \\
reference star $\langle (B-V) \rangle$ & & & $0.9\tablenotemark{a} $  \\
Absolute Parallax\tablenotemark{b}  & & &311.37 $\pm$ 0.11     mas \\
Relative Proper Motion & &   & $976.54 \pm 0.1$  mas y$^{-1}$  \\
 \indent in pos. angle & & & 269\fdg0  $\pm$ 0\fdg6   \\
\hline
{\it HIPPARCOS} Absolute Parallax& &&310.74 $\pm$ 0.85 mas\\
{\it HIPPARCOS} Proper Motion  &&&976.52 $\pm$ 1.9 mas y$^{-1}$ \\
 \indent in pos. angle &&& 271\fdg1 $\pm$ 3\fdg8 \\
\enddata
\tablenotetext{a}{Estimated from $VJHK$ photometry with $A_V = 0.0$. }
\tablenotetext{b}{Value from modeling RV and \HST and MAP astrometry 
simultaneously}
\end{deluxetable}
%\end{center}

\begin{center}
\begin{deluxetable}{ll}
\tablecaption{ Orbital Elements of \eps~Perturbation Due to \eps~ b  \label{tbl-ORB}}
\tablewidth{0in}
\tablehead{\colhead{Parameter} &  \colhead{Value}
}
\startdata
$\alpha_A$&1.88 $\pm$ 0.20 mas\\
$\alpha_Asini$ & 3.02e-3 $\pm$ 0.32e-3 AU \\
P& 2502 $\pm$ 10 d\\
P& 6.85 $\pm$ 0.03 yr\\
T$_0$ & 54207 $\pm$7 mJD \\
T$_0$ & 2007.29 $\pm$ 0.02 y\\
e& 0.702 $\pm$ 0.039 \\
i& 30\fdg1 $\pm$ 3\fdg8 \\
$\Omega$&74\arcdeg $\pm$ 7\arcdeg \\
$\omega$&47\arcdeg $\pm$ 3\arcdeg \\
$K_1$&18.5 $\pm$0.2 \ms \\
$M_*$ & 0.83$\pm$0.05$M_{\sun}$\\
\enddata
\end{deluxetable}
\end{center}

\begin{center}
\begin{deluxetable}{lr}
\tablecaption{\eps~b Parameters \label{tbl-massprm} }
\tablewidth{4in}
\tablehead{
\colhead{Parameter}    &
 \colhead{Value}
}
\startdata
{\it a} (AU) & 3.39 $\pm$ 0.36\\
$\Omega$' & 254\arcdeg\\
%$\omega$' & 63\arcdeg\\
$\omega$' & 47\arcdeg\\
Mass Function ($M_{\sun}$)& 5.9e-10 $\pm$ 1.0e-10 \\
\msini (\mjup)\tablenotemark{a}  &  0.78 $\pm$ 0.08\\
M (\mjup)\tablenotemark{b} & 1.55 $\pm$ 0.22 \\
M (\mjup)\tablenotemark{c} & 1.55 $\pm$0.24\\
\enddata
\tablenotetext{a}{derived from radial velocity alone}
\tablenotetext{b}{derived from radial velocity and astrometry, using Msini/sini}
\tablenotetext{c}{derived from radial velocity and astrometry, using
$m_2^3/(m_1 + m_2)^2 = a^3/P^2$; includes host star mass uncertainty.}
\end{deluxetable}
\end{center}

\clearpage

% And finally, we must deal with the figures.  There are three figures
% associated with this manuscript; two figures are Encapsulated
% PostScript (EPS) files.  The third figure is a grey scale figure that does
% not exist in EPS form.
%
% Authors have three options for including figure information within a
% manuscript.  Not all the options may be acceptable by the target Journal 
%- be
% sure to look at the appropriate submission instructions, electronic or
% otherwise.
%

%\clearpage

%\clearpage

%\clearpage

\clearpage

\clearpage

\clearpage

%\clearpage

%\clearpage
%\clearpage

\end{document}